# Optical cooling of dot-in-crystal halide perovskites: challenges of nonlinear exciton recombination

*Yasuhiro Yamada[1],\*, Takeru Oki[1], Takeshi Morita[1], Takumi Yamada[2], Mitsuki Fukuda[3], Shuhei Ichikawa[3], Kazunobu Kojima[3], and Yoshihiko Kanemitsu[2]*

1)Graduate School of Science, Chiba University, Inage, Chiba 263-8522, Japan

2)Institute for Chemical Research, Kyoto University, Uji, Kyoto 611-0011, Japan

3)Graduate School of Engineering, Osaka University, Suita, Osaka 565-0871, Japan

ABSTRACT

Highly efficient anti-Stokes (AS) photoluminescence (PL) is observed from halide perovskite quantum dots (QDs) due to their strong electron-phonon interactions. The AS PL is particularly intriguing as it suggests the potential for semiconductor optical cooling if the external quantum efficiency approaches 100%. However, the PL quantum efficiency in QDs is primarily dominated by multiparticle nonradiative Auger recombination processes under intense photoexcitation, which impose limits on the optical cooling gain. Here, we investigate the Auger recombination of dot-in-crystal perovskites. We quantitatively estimate the maximum optical cooling gain and the corresponding excitation intensity. We further conducted optical cooling experiments and demonstrate a maximum photo-cooling of approximately 9 K from room temperature. Additionally, we confirmed that increasing the excitation intensity leads to a transition from photo-cooling to photo-heating. These observations are consistent with our time-resolved measurements, offering insights into the potential and limitations of optical cooling in semiconductor QDs.

Solid-state optical cooling refers to a technique that cools materials using anti-Stokes (AS) photoluminescence (PL), where the emitted light has higher energy than the excitation light energy. If the external quantum efficiency reaches 100%, the material loses its internal energy and be cooled down through the process of light absorption followed by AS emission.[1,2] Since the first realization in 1995 with highly luminescent rare-earth-ion-doped glasses, solid-state optical cooling has been an area of active research.[3,4] Recently, there has also been growing interest in semiconductor-based optical cooling. The significant absorption coefficients of semiconductors and their advanced device fabrication technologies have the potential to pave the way for the development of compact and innovative cooling devices.

Semiconductor optical cooling has been explored in materials such as GaAs-based heterostructures and CdS thin films.[5-9] Recently, halide perovskites have also emerged as promising candidates for this application.[10-13] These perovskite semiconductors exhibit a large AS shift due to their strong electron-phonon interactions, a major advantage for optical cooling.[14-16] Additionally, their defect tolerance leads to small nonradiative recombination rate and high PL efficiency.[17,18] These properties make perovskite semiconductors ideal for optical cooling. The advent of such novel materials sheds new light on the research in semiconductor optical cooling.[19,20]

While there have been some reports on successful semiconductor optical cooling, their validity is still under debate, and the understanding of their physical mechanisms remains insufficient.[21, 22] We believe that advancing research in semiconductor optical cooling requires addressing two pivotal issues. The first is achieving near-unity PL efficiency. Unlike rare-earth elements with 4f electron orbitals that are less susceptible to external influences due to the limited spatial extension, nonradiative recombination processes are usually nonnegligible in semiconductors. Therefore, it is more challenging to realize semiconductors with near-unity PL efficiency. A promising solution is utilizing quantum dots (QDs), which have achieved nearly 100% PL efficiency in various semiconductors.[23] However, in semiconductor QDs, nonradiative recombination rates increase prominently under intense photoexcitation due to multiparticle Auger recombination, leading to reduced PL efficiency.[24-27] Thus, as the excitation light intensity increases, a transition from photo-cooling to photo-heating will occur. Understanding Auger recombination in the AS PL process is crucial for elucidating the possibilities and limitations of optical cooling.

The second point is the temperature estimation method. Because the photo-cooling power is extremely small, non-contact temperature measurement techniques are required. Previous studies have utilized PL thermometry that exploits the temperature dependence of PL peak energy and intensity.[8,11,28] However, continuous light exposure can alter the optical properties of the sample (e.g., defect formation or sample degradation), directly affecting the PL intensity and indirectly influencing the PL peak energy through reabsorption of PL.[29] Therefore, these conventional methods may not always be reliable.

In this study, we first investigate the recombination processes in perovskite QDs to elucidate the limitations of optical cooling. We focus on the stable dot-in-crystal structure of perovskites ($CsPbBr_3$ QDs in a $Cs_4PbBr_6$ host crystal, referred to as $CsPbBr_3/Cs_4PbBr_6$), evaluating the quantized Auger recombination process using time-resolved spectroscopy. Based on the experimental results, we estimate the maximum cooling gain and corresponding excitation intensity. Subsequently, we attempt optical cooling of $CsPbBr_3/Cs_4PbBr_6$ microcrystals using a novel PL thermometry method. We demonstrate maximum photo-cooling of approximately 9 K from room temperature. Furthermore, a crossover from photo-cooling to photo-heating, dependent on the excitation light intensity, is also observed. The results of this study provide significant insights into the optical cooling in semiconductor QDs.

We fabricated dot-in-crystal perovskite, CsPbBr$_3$/Cs$_4$PbBr$_6$, by solution process (Supporting Information Section 1). Compared to bare semiconductor QDs or those surrounded by organic ligands, the dot-in-crystal structure offers enhanced stability because it is protected by the robust host crystal.[18] The bandgap energy of CsPbBr$_3$ is around 2.4 eV, thus exhibiting green PL. The host crystal Cs$_4$PbBr$_6$ has a larger bandgap energy (~4 eV) and is transparent in the visible region.

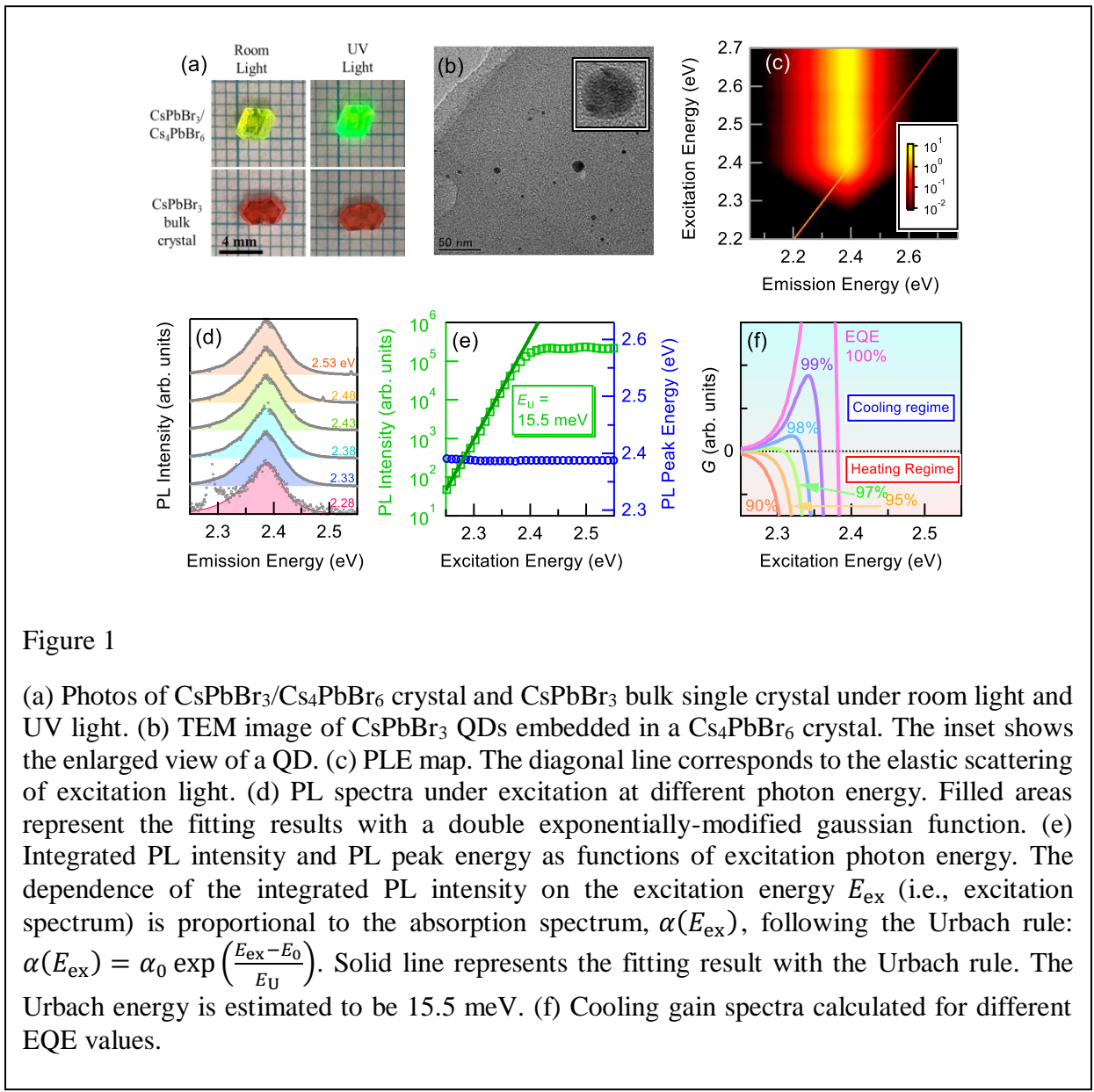

Figure 1

(a) Photos of CsPbBr$_3$/Cs$_4$PbBr$_6$ crystal and CsPbBr$_3$ bulk single crystal under room light and UV light. (b) TEM image of CsPbBr$_3$ QDs embedded in a Cs$_4$PbBr$_6$ crystal. The inset shows the enlarged view of a QD. (c) PLE map. The diagonal line corresponds to the elastic scattering of excitation light. (d) PL spectra under excitation at different photon energy. Filled areas represent the fitting results with a double exponentially-modified gaussian function. (e) Integrated PL intensity and PL peak energy as functions of excitation photon energy. The dependence of the integrated PL intensity on the excitation energy $E_{\text{ex}}$ (i.e., excitation spectrum) is proportional to the absorption spectrum, $\alpha(E_{\text{ex}})$, following the Urbach rule: $\alpha(E_{\text{ex}}) = \alpha_0 \exp\left(\frac{E_{\text{ex}} - E_0}{E_{\text{U}}}\right)$. Solid line represents the fitting result with the Urbach rule. The Urbach energy is estimated to be 15.5 meV. (f) Cooling gain spectra calculated for different EQE values.

The green color of CsPbBr$_3$/Cs$_4$PbBr$_6$ comes from the near-unity PL efficiency (Fig. 1(a)). Transmission electron microscope (TEM) observation reveals the spherical shape of CsPbBr$_3$ QDs with several nanometers in diameter (Figs. 1(b), S1-S3). The size of these QDs is comparable to the exciton Bohr radius,[14,30] suggesting the emergence of phenomena specific to QDs, such as quantum confinement effects and quantized Auger recombination.

To evaluate the AS PL characteristics of our CsPbBr$_3$/Cs$_4$PbBr$_6$ sample, we performed PL excitation (PLE) measurements. Figure 1(c) presents the PLE map for CsPbBr$_3$/Cs$_4$PbBr$_6$, where AS PL is observed below the diagonal line corresponding to the scattering of the excitation light. As illustrated in Figure 1(d), the PL spectrum shape is almost independent of the excitation photon energy, suggesting a negligible contribution from inhomogeneous broadening. As summarized in Figure 1(e), the PL peak energy remains unchanged even under below-bandgap excitation. This is in contrast to perovskite thin films and bulk crystals, where a redshift of the PL spectrum is observed at lower energy excitation.[15] This observed difference is attributable to the reduced PL reabsorption because of the relatively small photoabsorption of CsPbBr$_3$/Cs$_4$PbBr$_6$.[31] Considering that inhomogeneous broadening would decrease the AS shift (and therefore the cooling gain), CsPbBr$_3$/Cs$_4$PbBr$_6$ is an ideal material for optical cooling.

Furthermore, the low-energy tail of the excitation spectrum follows the Urbach rule over more than three orders of magnitude. The Urbach energy ($E_\mathrm{U}$) was determined to be 15.5 meV, which is consistent with or slightly lower than previous reports.[12] The Urbach energy can be influenced by contributions from intrinsic electron-phonon interactions and extrinsic inhomogeneities. However, as indicated by the PLE results, the contribution from inhomogeneities is almost negligible in this case. Therefore, the observed Urbach energy is likely to represent the contribution purely from electron-phonon interactions.

Based on these results, we can calculate the cooling gain, which is the net energy loss from the system per unit time. The dependence of the cooling gain per incident photon flux density on the excitation energy, or the cooling gain spectrum $G(E_{ex})$, is given by

$$G(E_{ex}) \propto \{\langle E_{PL}\rangle \eta_{ext} - E_{ex}\}\alpha(E_{ex}), \qquad (1)$$

where $\eta_{ext}$ is the external quantum efficiency (EQE) of PL.[32] $\langle E_{PL}\rangle$ represents the spectrally averaged PL energy. It is assumed here that the photoabsorption, $\alpha(E_{ex})$, is sufficiently small. The cooling gain spectrum calculated based on eq. (1) is presented in Fig.1(f). In the case of 100% EQE, optical cooling is possible with excitations below 2.38 eV, while higher energy excitations lead to photo-heating. As the EQE decreases, the cooling regime narrows, and at an EQE below 97%, optical cooling is hardly achievable at any excitation energy.

On the other hand, the EQE of our $CsPbBr_3/Cs_4PbBr_6$ samples gives the maximum efficiency of slightly above 80% (Fig. S4), which is below the cooling criteria. However, significant losses in PL efficiency are largely due to PL reabsorption, suggesting that higher efficiencies could be achieved with smaller, micrometer-sized crystals. This issue will be revisited in the section on cooling experiments.

If the PL efficiency is independent of the excitation light intensity, the cooling gain would increase proportionally with it. However, in semiconductors, nonradiative Auger recombination process becomes significant under high excitation densities. Auger recombination is particularly enhanced in QDs due to strong spatial confinement. This process reduces the PL quantum efficiency, potentially leading to photo-heating rather than photo-cooling under intense photoexcitation. Thus, it is essential to investigate the Auger processes and determine the intensity dependence of the cooling gain.

Figure 2(a) displays the PL relaxation dynamics at different excitation photon flux densities. The excitation energy was set at 2.75 eV, where Stokes PL is dominant. Under weak excitation, the PL decay profile is slightly non-exponential in shape, indicating the PL is contributed from an ensemble of QDs with slightly different exciton lifetimes. The effective lifetime $\tau_{\text{eff}}$ under weak excitation was determined to be 29 ns. As the excitation photon flux density increases, a faster relaxation component emerges. Figure 2(b) shows the dependence of $\tau_{\text{eff}}$ on the excitation photon

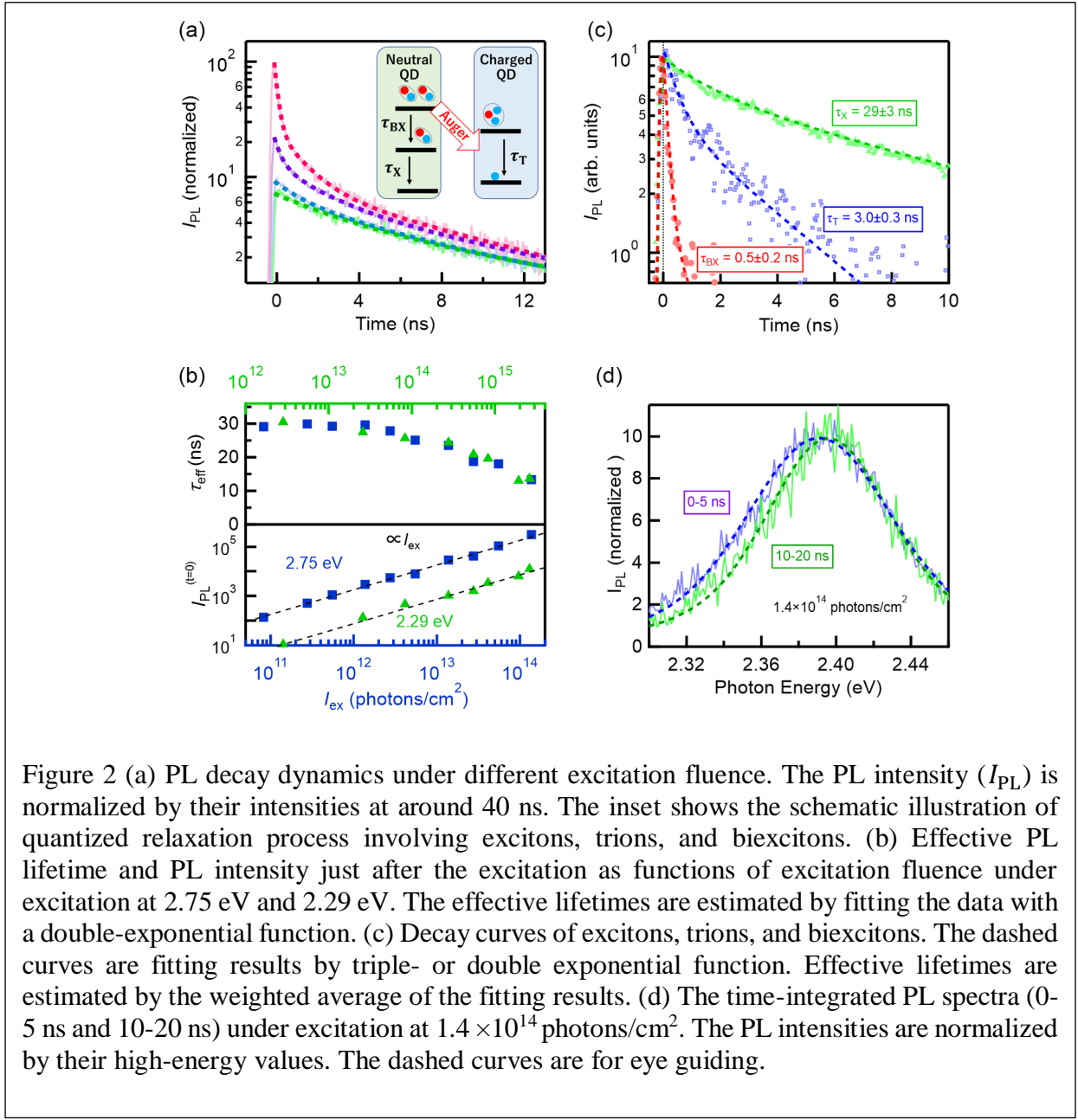

Figure 2 (a) PL decay dynamics under different excitation fluence. The PL intensity ($I_{\text{PL}}$) is normalized by their intensities at around 40 ns. The inset shows the schematic illustration of quantized relaxation process involving excitons, trions, and biexcitons. (b) Effective PL lifetime and PL intensity just after the excitation as functions of excitation fluence under excitation at 2.75 eV and 2.29 eV. The effective lifetimes are estimated by fitting the data with a double-exponential function. (c) Decay curves of excitons, trions, and biexcitons. The dashed curves are fitting results by triple- or double exponential function. Effective lifetimes are estimated by the weighted average of the fitting results. (d) The time-integrated PL spectra (0-5 ns and 10-20 ns) under excitation at $1.4 \times 10^{14}$ photons/cm². The PL intensities are normalized by their high-energy values. The dashed curves are for eye guiding.

flux density. This faster relaxation component is attributable to nonradiative Auger recombination. The same figure also displays the effective lifetime under excitation at 2.29 eV, a below-bandgap energy. The intensity dependence follows the same trend as that with 2.75 eV excitation, suggesting that Auger recombination is dominant under high-density excitation regardless of the excitation energy. In both cases, the initial PL intensity immediately after excitation linearly increases with the excitation density, ensuring a one-photon excitation process (Fig. 2(b)).

In semiconductor QDs, it is known that exciton recombination processes are quantize.[24-27] Particularly, in perovskite QDs, only three relaxation processes through excitons, trions (charged excitons), and biexcitons are observed due to the low degeneracy of the conduction and valence bands.[33,34] By taking the difference of the normalized PL relaxation profiles, we are able to extract the relaxation dynamics of the above-mentioned three components. (Fig. S5).[26,35] The effective lifetimes, as summarized in Figure 2(c), for excitons, trions, and biexcitons are $\tau_X$ = 29 ns, $\tau_T$ = 3.0 ns, and $\tau_{BX}$ = 0.5 ns, respectively. These lifetimes are consistent with those reported previously.[34] The PL spectrum also indicates the presence of trions and biexcitons. PL spectra within a few ns after excitation exhibits the redshift under intense photoexcitation (Fig. 2(c)), indicating the presence of a lower-energy component from biexcitons or trions.

Here we consider the PL efficiency of trions and biexcitons. Theory suggests that the radiative lifetimes of trions and biexcitons are 1/2 and 1/4 of the exciton radiative lifetime, respectively.[33,36] Assuming a PL efficiency of $\eta_X$ = 0.8 for excitons (Fig. S4), the estimated radiative lifetime of excitons, $\tau_X^{rad}$, is approximately 23 ns. Consequently, we estimate the radiative lifetimes of trions and biexcitons at $\tau_T^{rad} \approx$ 12 ns and $\tau_{BX}^{rad} \approx$ 6 ns, respectively. From these values, the PL efficiencies for trions and biexcitons are calculated to be $\eta_T$ = 0.26 and $\eta_{BX}$ = 0.09, respectively. Such notably low PL efficiencies indicate that their recombination processes are primarily dominated by nonradiative Auger recombination, which significantly reduces the PL efficiency and acts as a major impediment to optical cooling.

Based on the above results, we estimate the PL quantum efficiency and the cooling gain under continuous-wave excitation. As depicted in the inset of Fig. 2(a), we adopt a recombination model that involves excitons, trions, and biexcitons. Building on the previous work by Nakahara et al., we postulate that charged QDs are generated through Auger recombination.[37] Given these conditions, the PL quantum efficiency is described using the dimensionless coefficients $A$ and $C$ (Supporting Information Section 8). Consequently, the external quantum efficiency can be expressed by

$$\eta_{\text{ext}} = \frac{\eta_X + \eta_{BX}(\beta\tau_X) + \eta_T C(\beta\tau_X)^2}{1 + (1 + A)(\beta\tau_X) + C(\beta\tau_X)^2} \quad (2)$$

where $\beta$ represents the probability of a single QD being excited per unit time, which is assumed to be proportional to the excitation intensity (i.e., the effects of photobleaching are considered negligible). We fitted the excitation-intensity dependence of the PL efficiency of our samples using Eq. (2) and found that the coefficients $A$ and $C$ are negligibly small (Fig. S7). The cooling gain per unit time for a single QD, $g$, can be calculated from

$$g = \beta(\eta_{\text{ext}}\langle E_{\text{PL}}\rangle - E_{\text{ex}}). \quad (3)$$

Figure 3 shows the calculated $\eta_{\text{ext}}$ and $g$ as functions of excitation intensity. Here, the excitation photon energy is set at 2.33 eV, which is actually used in the subsequent optical cooling experiments. For simplicity, the coefficients $A$ and $C$ are assumed to be zero. Additionally, in the most favorable case, we assumed the PL efficiency of the single exciton state ($= \eta_X$) to be 100%. The cooling gain reaches its maximum ($g_{\max}$ = 2.1 fW per QD) at an intensity corresponding to $\beta\tau_X$ = 0.014. As the excitation intensity is increased beyond this level, the cooling gain rapidly

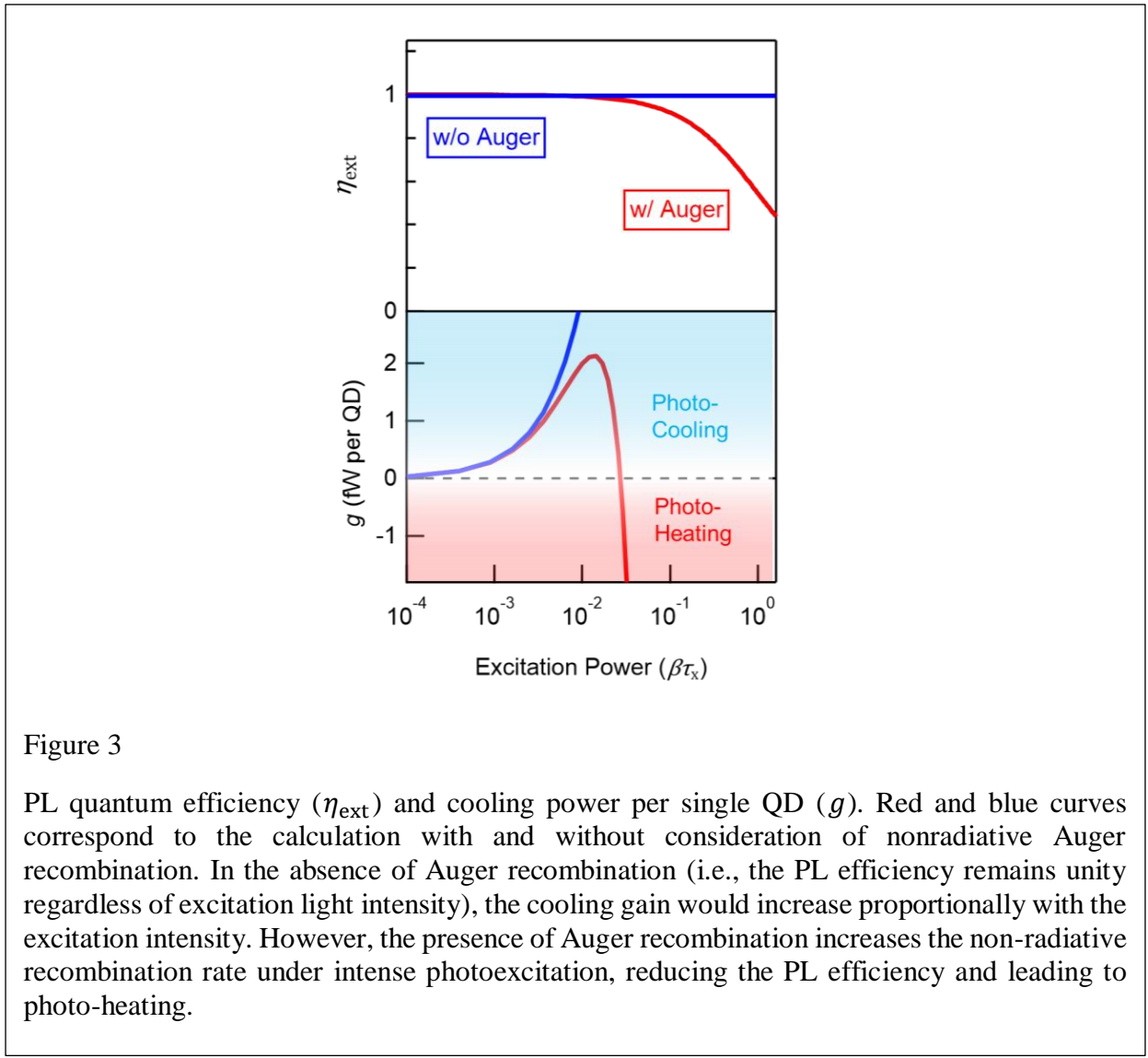

Figure 3

PL quantum efficiency ($\eta_{\text{ext}}$) and cooling power per single QD ($g$). Red and blue curves correspond to the calculation with and without consideration of nonradiative Auger recombination. In the absence of Auger recombination (i.e., the PL efficiency remains unity regardless of excitation light intensity), the cooling gain would increase proportionally with the excitation intensity. However, the presence of Auger recombination increases the non-radiative recombination rate under intense photoexcitation, reducing the PL efficiency and leading to photo-heating.

decreases, and a transition from photo-cooling to photo-heating is found. This result implies that there is an optimal excitation density for optical cooling.

The above estimation provides key insights into the appropriate excitation light intensity for designing optical cooling experiments. From rough estimation, we calculate the optimal excitation light intensity as approximately ~ kW/cm$^2$ (Supporting Information Section 5). This excitation intensity is easily achievable. Therefore, it is crucial to recognize that in optical cooling experiments, excitation light intensity exceeding this threshold may actually inhibit the intended cooling effects.

Additionally, it is also meaningful to estimate the lower limit of the cooling temperature that can be achieved with the calculated cooling power of 2.1 fW per QD. Simple calculations that consider only thermal radiation indicate that the limit of optical cooling for a thermally-isolated single QD is approximately 1 K from 300 K (Supporting Information Section 9). This estimate implies that cooling a single semiconductor QD by several kelvins is fundamentally impossible. In contrast, our target material $CsPbBr_3/Cs_4PbBr_6$, which contains numerous QDs, can be cooled down to about 10 K. Such a level of temperature change would be experimentally detectable.

Based on the insights gained from PLE and time-resolved spectroscopy, we conducted optical cooling experiments with $CsPbBr_3/Cs_4PbBr_6$. First, we need to address the challenges associated with temperature estimation. Here we utilize the high-energy exponential tail of PL spectra that is

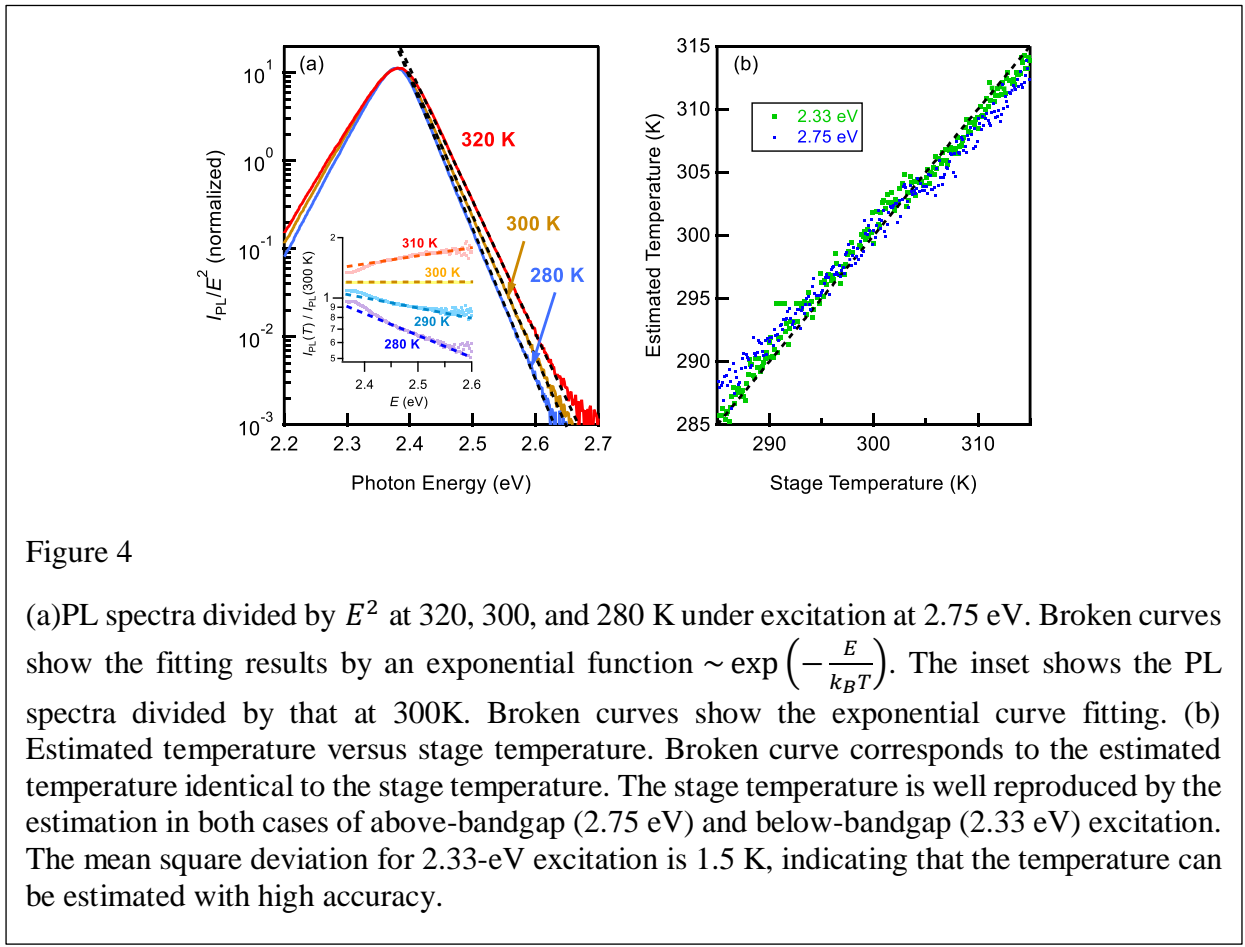

Figure 4

(a) PL spectra divided by $E^2$ at 320, 300, and 280 K under excitation at 2.75 eV. Broken curves show the fitting results by an exponential function $\sim \exp\left(-\frac{E}{k_B T}\right)$. The inset shows the PL spectra divided by that at 300K. Broken curves show the exponential curve fitting. (b) Estimated temperature versus stage temperature. Broken curve corresponds to the estimated temperature identical to the stage temperature. The stage temperature is well reproduced by the estimation in both cases of above-bandgap (2.75 eV) and below-bandgap (2.33 eV) excitation. The mean square deviation for 2.33-eV excitation is 1.5 K, indicating that the temperature can be estimated with high accuracy.

dependent on temperature (Supporting Information Section 3).[38] To examine this method, we estimated the temperature of a CsPbBr$_3$/Cs$_4$PbBr$_6$ sample placed on temperature-controlled metal stage with sufficient thermal contact. Figure 4(a) shows the PL spectra divided by $E^2$ under excitation at 2.75 eV at different temperatures. With decreasing temperature, the PL spectrum narrows, and the high-energy exponential tail becomes steeper. We further enhanced the accuracy of detecting temperature changes by normalizing the PL spectrum with that at a specific reference temperature and achieved highly accurate temperature estimation. As shown in Fig 4(b), the estimated results closely replicate the stage temperature. Furthermore, we have verified that this method yields satisfactory results even for excitation at 2.33 eV, where AS PL is dominant (Fig. 4(b)).

Building on the results above, we applied our PL thermometry technique to demonstrate optical cooling. For this purpose, we utilized CsPbBr$_3$/Cs$_4$PbBr$_6$ microparticles obtained by grinding the bulk sample. We chose this approach to enhance the light out-coupling, thereby improving the external PL quantum efficiency. In fact, by reducing the particle size, we observed an increase in external PL efficiency, likely due to reduced PL re-absorption (Fig. S4). Additionally, if there are spatial fluctuations in the PL efficiency, it might be possible to selectively utilize the areas with higher efficiency.

We selectively photoexcited a highly luminescent microparticle, which is placed on a mica substrate in a vacuum chamber (Fig. 5(a)). Here, we present the optical cooling results from the sample that showed the most significant temperature reduction. Figure 5(b) displays the high-energy side of the PL spectra immediately after photoexcitation and 10 minutes later. The spectrum after 10 minutes exhibits a slightly steeper shape compared to the one immediately after excitation, suggesting a decrease in temperature. This decrease in temperature becomes more evident when

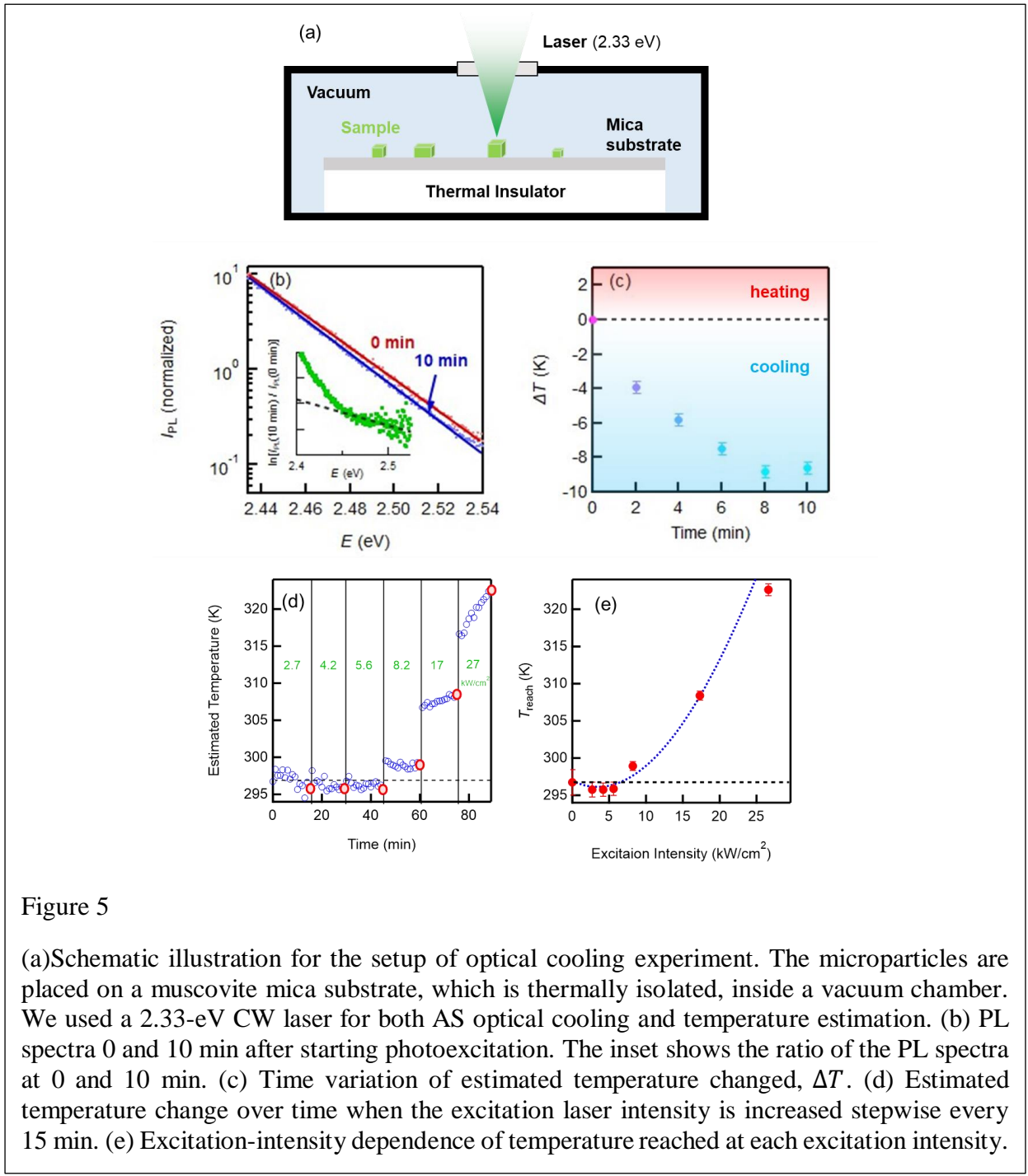

Figure 5

(a) Schematic illustration for the setup of optical cooling experiment. The microparticles are placed on a muscovite mica substrate, which is thermally isolated, inside a vacuum chamber. We used a 2.33-eV CW laser for both AS optical cooling and temperature estimation. (b) PL spectra 0 and 10 min after starting photoexcitation. The inset shows the ratio of the PL spectra at 0 and 10 min. (c) Time variation of estimated temperature changed, $\Delta T$. (d) Estimated temperature change over time when the excitation laser intensity is increased stepwise every 15 min. (e) Excitation-intensity dependence of temperature reached at each excitation intensity.

the spectrum is normalized (see the inset of Fig. 5(b)). Figure 5(c) plots the estimated temperature against elapsed time. Thus, cooling of about 9 K was observed over 10 minutes. This value is close to the limit of optical cooling (Supporting Information Section 9).

Next, to investigate the impact of nonradiative Auger recombination on optical cooling, we conducted measurements with varying excitation intensity. Figure 5(d) shows the changes in estimated temperature when the excitation intensity was varied from 2.7 kW/cm² to 27 kW/cm². While slight optical cooling was observed at lower excitation intensities, a transition from cooling to heating occurs as increasing excitation intensity. This transition is attributable to the photo-heating effects by nonradiative Auger recombination. The relationship between the reached temperature and excitation light intensity at each excitation intensity is plotted in Fig. 5(e). The lowest temperature was achieved at an excitation intensity of around a few kW/cm², aligning well with our estimation (~ kW/cm², Supporting Information Section 5). Assuming a proportional relationship between cooling gain and the achieved temperature, we fitted the data using Eqs. (2) and (3), represented by the blue curve in the figure, which generally reproduces the temperature changes. These results clearly indicate that Auger recombination defines the limit of optical cooling.

In conclusion, we discussed the limitation of optical cooling in semiconductor QDs, determined by Auger recombination. The increase in the nonradiative recombination rate under high-density excitation is a clear inhibitory factor for optical cooling. Based on the relaxation time constants of excitons, trions, and biexcitons estimated from time-resolved PL data, we quantitatively calculated the excitation-intensity dependence of optical cooling gain. From the maximum cooling gain per QD, we determined the lowest achievable temperature for an isolated QD. We revealed that cooling of only about 1 K is possible with a single QD. However, optical cooling to lower temperatures is achievable in QD ensembles. In fact, we demonstrated evidence of optical cooling at weak excitation in a CsPbBr$_3$ QD ensemble embedded in a host crystal, observing cooling of 9 K. The excitation-intensity dependence of the cooling temperature showed good agreement with our calculations based on Auger recombination, clearly delineating the limits of optical cooling in semiconductor QDs.

Our research provides important insights for the design of semiconductor optical cooling devices. When using QDs, it is essential to suppress the Auger recombination rate. This might be achieved by appropriately selecting the material surrounding the QDs, as demonstrated in core-shell structures.[39,40] As reported recently, the biexciton Auger rate of halide perovskites deviates from the volume scaling law in the weakly confined regime.[41,42] This would also be helpful in increasing $\eta_{BX}$. Furthermore, as the cooling gain per QD is limited, increasing the density of QDs is essential for maximize the cooling power. Even in these cases, the influence of Auger recombination is inevitable, necessitating the development of methods for its assessment and suppression.

ASSOCIATED CONTENT

**Supporting Information**.

The following files are available free of charge.

Material growth, experimental setup, supplementary experimental results (PDF)

AUTHOR INFORMATION

**Corresponding Author**

*Yasuhiro Yamada; Graduate School of Science, Chiba University, Inage, Chiba 263-8522, Japan, yasuyamada@chiba-u.jp

**Author Contributions**

The manuscript was written through contributions of all authors. All authors have given approval to the final version of the manuscript.

**Notes**

The authors declare no competing financial interest.

ACKNOWLEDGMENT

The authors thank Kenya Suzuki for the help of optical experiments. Part of this work was supported by the Canon Foundation, the International Collaborative Research Program of Institute for Chemical Research, Kyoto University (Grant No. 2023-21), JST-CREST (Grant No. JPMJCR21B4), and KAKENHI (Grant No. JPJP19H05465).